# Hydrogen Bond-Driven Interactions Between Chitosan and Biobased Surfactants: A Study of Bulk Behavior and Surface Adsorption

Ana Puente-Santamaría,[a] Josselyn N. Molina-Basurto,[a] Eva Gerardin,[c] Francisco Ortega,[ab] Ramón G. Rubio,[ab] and Eduardo Guzmán[ab*]

[a.] Departamento de Química Física, Universidad Complutense de Madrid. Plaza de las Ciencias 2- Ciudad Universitaria. 28040-Madrid (Spain)

[b.] Instituto Pluridisciplinar, Universidad Complutense de Madrid. Paseo de Juan XXIII 1. 28040- Madrid (Spain)

[c.] Ecole Nationale Supérieure de Chimie de Lille (ENSCL). Cité Scientifique - Bât. C7. Avenue Mendeleïev - CS 90108. 59652-Villeneuve D'Ascq Cedex (France)

* To whom correspondence should be addressed: eduardogs@quim.ucm.es

**Abstract**

This study explores the hydrogen bond-mediated association between chitosan (CHI) and alkyl polyglucoside (APG), a bio-based surfactant, in acidic conditions with varying ionic strengths. Unlike conventional polyelectrolyte-surfactant interactions that depend on electrostatic forces, the association in this system relies purely on non-ionic interactions. Using UV-visible spectroscopy, phase diagrams, and quartz crystal microbalance with dissipation monitoring (QCM-D), the bulk phase behavior and adsorption characteristics of CHI-APG mixtures on negatively charged surfaces was studied. Results demonstrate that APG concentration controls the phase behavior, with moderate levels inducing coacervate formation, while higher ionic strengths promote this coacervation through enhanced hydrogen bonding interactions. This shift leads to the formation of a phase-separated morphology, with micron-sized coacervate droplets observable in solution. Zeta potential measurements suggest that these droplets adopt a core-shell structure, characterized by a hydrophobic core due to the surfactant's alkyl chains and a hydrophilic shell formed by chitosan. Additionally, the coacervation process significantly enhances the adsorption of CHI-APG complexes onto solid substrates, a feature with potential applications in targeted delivery and controlled release systems. Overall, this study provides critical insights into the design of bio-based, sustainable formulations and expands the understanding of hydrogen bond-driven, non-electrostatic coacervation, relevant for applications in cosmetics, biomedical coatings, and environmentally friendly materials.

## 1. Introduction

Polyelectrolyte-surfactant mixtures have garnered significant attention due to their versatile applications and intricate physicochemical properties. These mixtures play an indispensable role across industries such as food technology, personal care, drug delivery, and enhanced oil recovery.[1–3] This is partly due to the intricate phase behavior of such mixtures, including the ability to undergo liquid-liquid (coacervation) and liquid-solid (precipitation or sedimentation) phase separations under specific conditions.[4] The latter is very common in oppositely charged polyelectrolyte-surfactant systems, occurring when there is a matching between the number of charged monomers and surfactant molecules in solution. However, coacervation does not require the participation of electrostatic interactions as demonstrated by Jing et al.[5] in mixtures of poly(ethyleneglycol) and poly-oxometalates. Coacervates form a unique class of supramolecular complexes that form in aqueous environments. Although the individual components can be highly soluble in water, when combined at certain concentrations and temperatures, the solution spontaneously separates into two distinct liquid phases. These structures are commonly found in biological systems, including living cells (compartmentalization, and membraneless organelles), and in marine organisms like mussels and sandworms.[6] Moreover, they are related to the origin of life.[7]

Recent theoretical advances have expanded the classical view related to the driving forces of the coacervation by demonstrating that many-body dipolar correlations and solvent polarizability can induce self-coacervation even in the absence of dominant ionic interactions [8–10]. These studies highlight that subtle solvent–polymer interactions and the entropy gain associated with the release of hydration water can be key drivers of phase separation. Despite the pivotal role of water in coacervation, little research has been focused on biomimetic complex coacervation mediated directly by non-electrostatic interactions involving water, i.e., coacervation driven by the entropy gain associated with the release of hydration water. Moreover, emerging evidence from studies of local solvent phase transitions [11] and cosolvent-induced polymer swelling and collapse [12] supports a model in which non-ionic interactions and hydration dynamics critically influence the overall phase behavior of polymer systems. In this context, the present study explores the bulk association of chitosan (CHI), a biopolymer derived from chitin, and alkyl polyglucoside surfactant (APG), a bio-based surfactant. By investigating their behavior at a pH of 4.5 and varying ionic strengths, we highlight the association mechanisms that occur in the absence of ionic binding between the polyelectrolyte and the surfactant, a novel approach in contrast to most existing studies [13,14]. Furthermore, insights from recent work on dipolar polymer gels [15] indicate that leveraging non-ionic, dipolar interactions can lead to alternative self-assembly pathways and enhanced material functionalities, thereby offering fresh perspectives on the design of advanced bio-based systems.

The phase behavior of CHI-APG mixtures and their ability to undergo coacervation may offer new insights into the fundamental principles governing the behavior of bio-based polyelectrolyte-surfactant systems.[16,17] Understanding this behavior is crucial for optimizing industrial applications, improving environmental sustainability, and driving future innovation.[18,19]. In fact, chitosan, a biobased polysaccharide, has gained significant attention in biomedical and pharmaceutical applications due to its exceptional biocompatibility, biodegradability, and versatile functional properties. Chitosan has been widely utilized in drug delivery systems, offering a promising platform for the controlled release of therapeutic agents [20,21]. Its adaptability allows for the formulation of various delivery mechanisms, including films, with particular relevance in wound-healing applications, gels, or nanoparticles. Beyond drug delivery, chitosan also plays a crucial role in tissue engineering, where it serves as a scaffold to support cell adhesion, proliferation, and tissue regeneration [22]. Furthermore, its bioactive properties make it a valuable component in cosmetic formulations, where it functions as a skin-conditioning agent, antimicrobial additive, or film-forming ingredient [23].

## 2. Materials and Methods

### 2.1. Chemicals

Alkyl polyglucoside (APG), commercially available as Oramix GC-110, was obtained from Safic-Alcan (Barcelona, Spain). Oramix GC-110 consists of an aqueous solution (60% in weight of active matter) of an equal molar mixture (0.5 fraction) of caprylyl glucoside and capryl glucoside. Chitosan (CHI), with a molecular weight ranging from 100 to 300 kDa and a degree of deacetylation (DDA) of about 90%, was supplied by Thermo Fisher Scientific (Waltham, MA, USA). Figure 1 shows the molecular formula of APG and CHI.

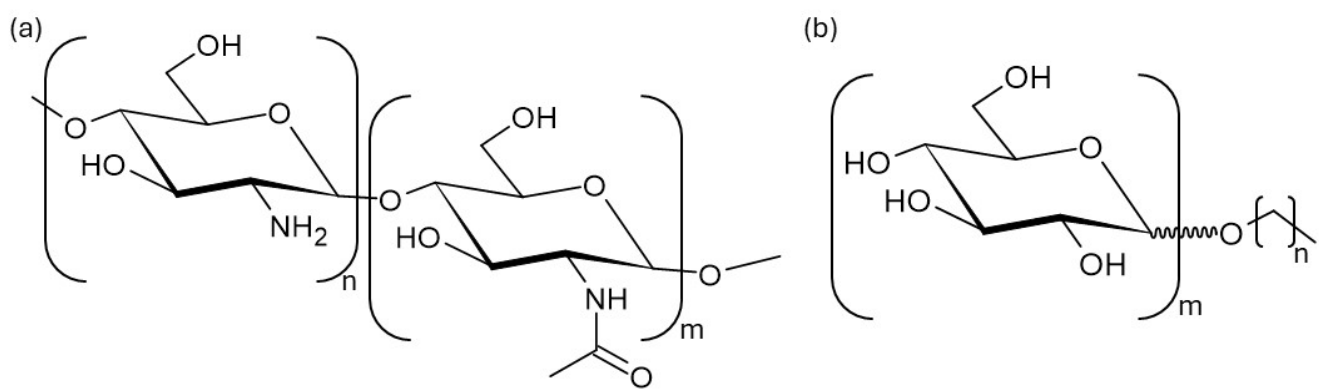

**Figure 1.** Molecular structures for chitosan (a) and APG (b). In panel (a), n y m indicate the deacetylated and acetylated monomers, respectively. It is worth remembering that deacetylated monomers correspond to 90% of the monomers in chitosan chains, whereas acetylated monomers only represent 10% of the monomers in chitosan chains. In the case of APG (panel (b)), m assumes a value of 1.5 as the used surfactant is an equimolar mixture containing surfactant molecules with one and two sugar rings as polar heads, and n assumes values of 7 and 9, as the surfactant mixture contains the same number of molecules having as hydrophobic tail octyl and decyl alkyl chains.

Glacial acetic acid (99.5% purity) and NaCl (99.95% purity), both obtained from Merck (Darmstadt, Germany), were used to adjust the pH and ionic strength of the solutions (stock chitosan and APG mixtures, and polyelectrolyte-surfactant mixtures).

Ultrapure Milli-Q grade deionized water, with a resistivity exceeding 18 MΩ·cm and a total organic content (TOC) below 6 ppm, was utilized. This high level of purity was achieved using an AquaMAX™-Ultra 370

Series multi-cartridge purification system (Young Lin Instrument Co., Ltd., Gyeonggi-do, Republic of Korea). All experiments were performed at 25ºC.

**2.2. Solution preparation**

The preparation of chitosan solutions and chitosan-APG mixtures were done by protocols adapted from those reported in literature.[24,25] In the following, these protocols are briefly summarized.

Chitosan stock solutions with a concentration of 10 mM relative to the fraction of deacetylated monomers were prepared by weighing the appropriate amounts of chitosan and NaCl and pouring them into a 50 mL flask. The flask was then partially filled with water, followed by the addition of 100 μL glacial acetic acid to lower the pH. The mixture was left overnight under mild stirring (1000 rpm) to ensure complete dissolution of the chitosan in the aqueous medium. The pH of the solution was then carefully adjusted to 4.5 by the gradual addition of 0.01 mM sodium hydroxide. The final solution was obtained by adding the required volume of a diluted aqueous acetic acid solution at pH 4.5 until a final volume of 50 mL was reached.

APG stock solutions with a concentration of 200 mM were prepared by weighing and adding the required amounts of commercial APG and NaCl to a 100 mL flask. A diluted aqueous acetic acid solution at pH 4.5 was then added to the flask to obtain the final solution. In some cases, diluted APG stock solutions were needed. These were prepared by dilution of a concentrated stock solution with a diluted aqueous acetic acid solution at pH 4.5 and fixed NaCl concentration.

The preparation of the mixtures of chitosan and APG requires the sequential addition of the required volumes of chitosan and APG stock solution into a 25 mL flask, followed by the dilution with a diluted aqueous acetic acid solution at pH 4.5 and fixed NaCl concentration. To ensure the homogenization of the mixtures, they were subjected to mild stirring (1000 rpm) for 24 hours. It is worth noting that the aging of the chitosan-APG mixtures does not result in any macroscopic change in the mixture.

It should be noted that as acetic acid is not a buffer, the pH was evaluated both after the preparation and immediately prior to sample use, ensuring that all measurements were consistently taken at a pH of 4.5.

**2.2. Experimental Methods**

A quartz crystal microbalance with dissipation monitoring (QCM-D), model Explorer from Qsense (Gothenburg, Sweden), equipped with gold-coated AT-cut quartz crystals, was utilized to examine the adsorption of polyelectrolyte-surfactant layers on negatively charged surfaces. The quartz sensors were prepared by first cleaning them with a Piranha solution (a 70% sulfuric acid/30% hydrogen peroxide mixture) for 30 minutes, followed by a thorough rinse with Milli-Q water. After cleaning, a self-assembled monolayer of 3-mercapto-1-propane sulfonic acid (CAS No. 49594-30-1) was formed on the sensor surfaces. This monolayer imparts a permanent negative charge to the gold substrate. The QCM-D technique measures the impedance spectrum of the quartz crystal at its fundamental frequency ($f_0$=5 MHz) and at odd harmonics, up to the 13$^{th}$ overtone (central frequency $f_{11}$=65 MHz). The obtained impedance spectra were analyzed using a single-layer model, based on the methodology proposed by Voinova et al.[26] This model correlates changes in resonance frequency ($\Delta f$) and dissipation ($\Delta D$) across different overtones with the physical properties of the adsorbed layers, such as thickness, density, elasticity, and viscosity. In this study, the viscoelastic nature of the adsorbed polymer film was incorporated into the QCM-D modeling, which is essential for accurately capturing the behavior of the system. Unlike rigid boundary conditions, the viscoelastic approach accounts for the heterogeneous adsorption typical of individual polymer layers. The high degree of hydration often found in polyelectrolyte layers adds a dynamic component to the system, which can only be properly modeled by incorporating the viscous response of the film. This method leads to a more precise representation of the mechanical properties and interactions between the film and the substrate, thereby offering a deeper understanding of its performance and stability across various applications.[27–30] For more comprehensive details on the data analysis procedure, please refer to reference.[31]

The effective charge density of the polyelectrolyte-surfactant aggregates dispersed in the aqueous phase may be inferred from measurements of the electrophoretic mobility, ue, obtained by Laser Doppler velocimetry measurements with a Nanosizer ZS (Malvern instruments, Malvern, UK). Electrophoretic mobility is directly proportional to the zeta potential, ζ, by the Henry's equation.[32]

UV-visible spectroscopy measurements were performed using an UV-visible spectrophotometer model V-730 (Jasco Inc. Ltd., Tokyo, Japan). This spectrophotometer allows obtaining the absorbance ($Abs$) in the wavelength range 190-600 nm. The same spectrophotometer was used for determining the turbidity of the samples at a fixed wavelength (450 nm), far from the absorption bands appearing in the system. Thus, it is possible to define the turbidity according to the following expression $Turb = 1 - T$, with T being the transmittance defined as $T = 10^{-Abs}$.

Centrifugation experiments at 14000 rpm for 1 hour were performed using a Microfuge® 18 Centrifuge (Beckman Coulter S.L.U., Indianapolis, IN, USA).

A Nikon Eclipse 80i microscope, which was fitted with a 50×, 0.55 NA LU Plan objective (Nikon, Japan) and connected to an ORCA-Flash4.0 V3 CMOS camera (model C13440-20CU, Hamamatsu Photonics K.K., Japan) was used to visualize the microscopic aspect of solutions. For this purpose, images with a resolution of 2048 × 2048 pixels, covering an area of 130 × 130 µm², were captured at a frame rate of 30 fps.

## 3. Results

### 3.1. Phase behavior of Chitosan-APG mixtures

The study of the phase diagram of the polyelectrolyte-surfactant system is of a paramount importance on the understanding of the association between the chitosan and the APG. Turbidity ($\tau$), obtained from the absorbance of the samples measured at a fixed wavelength of 450 nm where none of the components absorb, provides a simplified overview on the phase behavior of CHI-APG mixtures with a fixed CHI concentration (1 mM, referred to the concentration of ionizable monomers) and varying the surfactant ($c_{APG}$) and NaCl concentrations ($c_{NaCl}$). Figure 2a shows the dependence of the turbidity ($Turb$), derived from the absorbance of the sample measured at 450 nm

(see Section 2.2), on the surfactant concentration for mixtures with different ionic strengths (in the range 0-150 mM) and a fixed chitosan concentration.

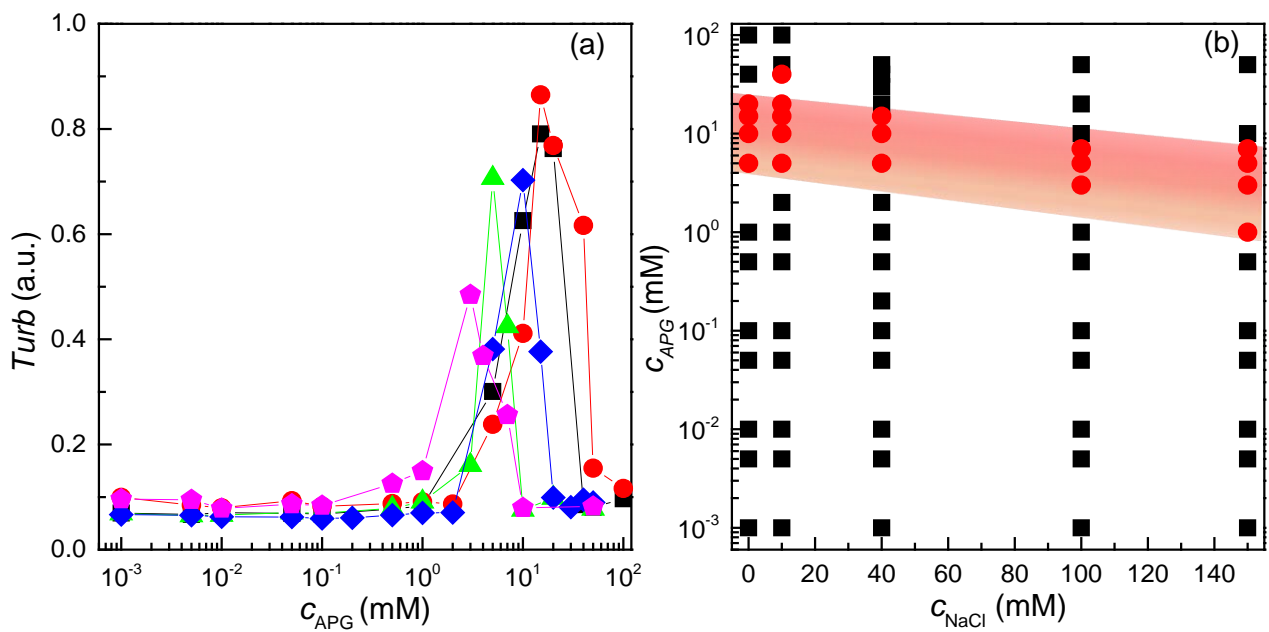

**Figure 2. (a)** Dependence of turbidity, determined at 450 nm, on APG concentration for CHI-APG mixtures, with fixed 1 mM chitosan concentration, prepared at different NaCl concentrations ($c_{NaCl}$). Different NaCl concentrations are represented by different coloured symbols. 0 mM (■), 10 mM (●), 40 mM (◆), 100 mM (▲) and 150 mM (⬟). The lines are guides for the eyes. **(b)** Phase diagram, represented as a $c_{APG}$ vs. $c_{NaCl}$ compositional map for CHI-APG mixtures. (■) and (●) correspond to transparent and turbid samples, respectively. The red shaded region represents the coacervation region.

The results show that independently of the ionic strength, the turbidity of the mixtures show a similar dependence on the APG concentration. At low APG concentration, the samples remain transparent, and the turbidity is close to the one of chitosan solutions at the same ionic strength. The increase of the APG concentration, below a threshold value, dependent on the ionic strength of the solutions, leads to an increase in sample turbidity reaching a maximum value. However, the samples does not show any signature of macroscopic phase separation like the one occurring in oppositely charged polyelectrolyte-surfactant systems, when a perfect matching between the number of binding sites in the polyelectrolyte chain and the number of bound surfactant molecules to the chain.[3] Further increases in APG concentration beyond the maximum lead to a decrease in the turbidity, corresponding to a new region where the samples present a transparent aspect similar to that of chitosan solutions as shown Figure 3.

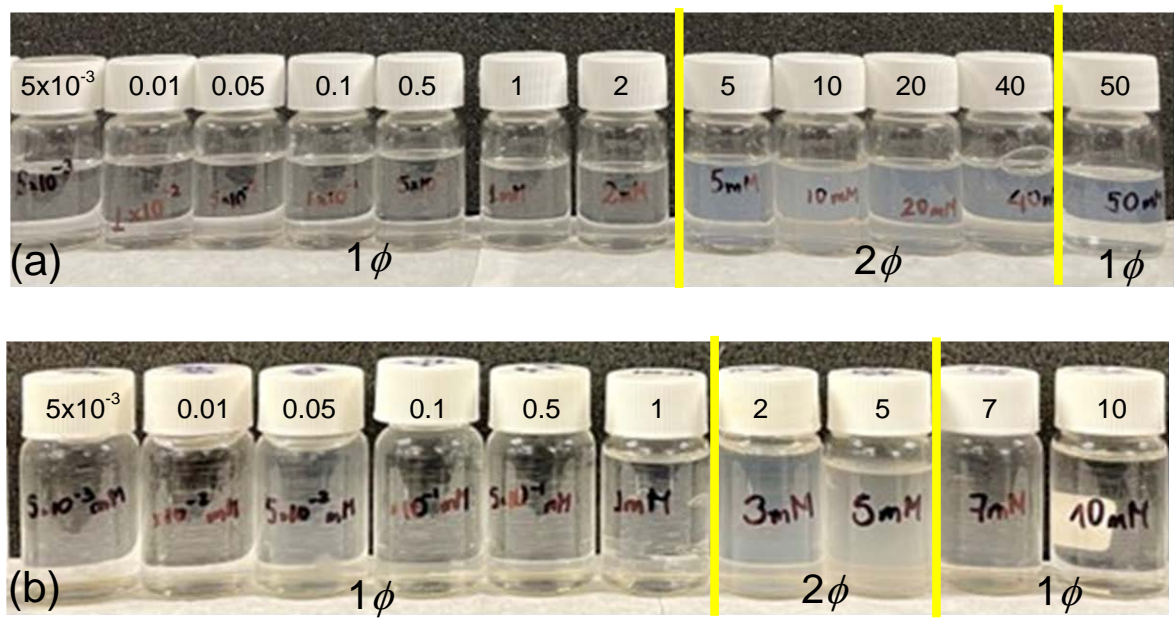

**Figure 3**. Set of images showing CHI-APG mixtures with a fixed chitosan concentration (1 mM) and increasing APG concentrations at two different NaCl concentrations ($c_{NaCl}$): $c_{NaCl}$ =10 mM **(a)** and 100 mM **(b)**. The numbers in the cap of the vials indicate the APG concentration in the mixture (in mM units) and the vertical lines indicate the separation between single phase solutions (1$\phi$) and biphasic mixtures (2$\phi$).

The above results suggests that even in the absence of electrostatic binding, it is possible to modulate the phase behavior of the polyelectrolyte-surfactant mixtures as a result of the association process, resulting in the appearance of different phase transitions as the CHI:APG ratio is changed. Figure 2b shows the phase diagram obtained for the system where the region corresponding to turbid samples is shaded. Despite the absence of any macroscopic evidence, it will be shown that turbid samples correspond to coacervates. Therefore, the increase in the turbidity of the sample may be the result of a phase transition from single-phase solutions (1$\phi$) to biphasic systems (2$\phi$).

In a first approximation, the phase behavior of the CHI-APG system is independent of ionic strength. However, a more detailed analysis reveals that an increase in ionic strength, increasing NaCl concentration ($c_{NaCl}$), has the effect of shifting the coacervation region towards compositions with lower APG concentrations ($c_{APG}$). The addition of NaCl effectively screens the charges on polyelectrolyte chains, reducing the electrostatic repulsive forces between them. This decrease in repulsion creates a "salting-out" effect, which disrupts water-chitosan and water-APG hydrogen bonds. As a result, this condition favors direct hydrogen bonding between chitosan and surfactant molecules, facilitating closer packing and aggregation of chitosan and APG. This enhanced association lowers the concentration of surfactant required to initiate coacervation, as the increased interactions lead to the formation and separation of coacervate droplets from the solution at lower APG levels. This aligns with both Density Functional Theory and DLVO predictions [33,34] and highlights the intricate balance between ionic strength and interactions in determining the phase behavior of the system.[35,36] In order to verify the existence of a true liquid-liquid phase separation, the samples were subjected to centrifugation. The results of this test can be observed in Figure 4 which shows the macroscopic aspect of the samples before and after centrifugation.

Initially, the dispersions exhibit a noticeable turbidity (Figures 3a and 3c), indicative of a heterogeneous system with dispersed aggregates. Upon centrifugation, a distinct white sediment forms at the bottom of the vials (Figures 3b and 3d), which suggests phase separation and the settling of a denser component. This behavior is consistent with the formation of complex coacervates, as reported by Bago-Rodríguez et al.[37] for interpolyelectrolyte complexes. Figure 4e illustrates this coacervation process, revealing numerous micron-sized droplets dispersed in the continuous medium. These droplets, approximately a few microns in diameter, are characteristic of complex coacervate systems, where microdroplet formation results from phase separation process.

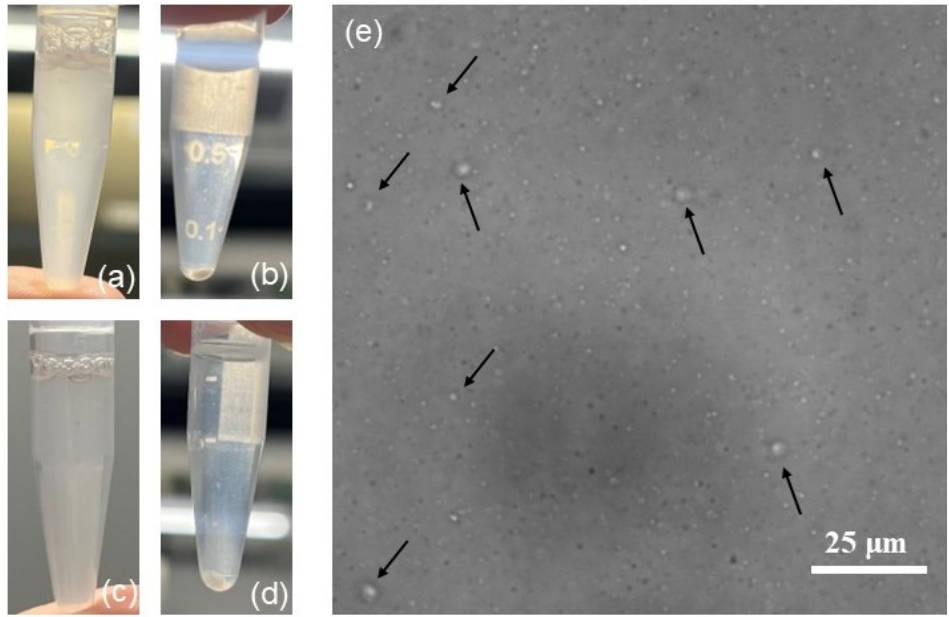

**Figure 4. (a-d)** Set of images corresponding to two CHI-APG mixtures with fixed chitosan (1 mM) in pure water and two different APG concentrations as prepared **(a,c)** and after centrifugation during 1 hour at 14000 rpm **(b,d)**. Panels **(a)** and **(b)** correspond to sample with $c_{APG}$=10 mM and panels **(c)** and **(d)** correspond to sample with $c_{APG}$=15 mM. **(e)** Optical microscope image for a CHI-APG mixtures with fixed chitosan (1 mM) in pure water and $c_{APG}$=15 mM. The arrows indicate for clarity the position of some of the coacervate droplets.

The differences on the macroscopic character between the two phases was confirmed by comparing the images of the mixtures as prepared and that corresponding to the supernatant remaining after centrifugation. Figure 5a presents an image depicting a series of samples from the coacervation region at a NaCl concentration of 10 mM. The image illustrates both the freshly prepared samples and the supernatant remaining after centrifugation. Prior to centrifugation, the samples exhibit a turbid appearance, which disappears in the supernatant obtained following centrifugation. This observation confirms that the turbidity is indicative of a multiphasic system, where a dense phase—dispersed within the aqueous environment—sediments during centrifugation. The presence of this second phase is further supported by the UV-visible spectra, represented as the absorbance (*Abs*) dependence on the wavelength, for both the original samples and the supernatant obtained following centrifugation, shown in Figures 4b and 4c.

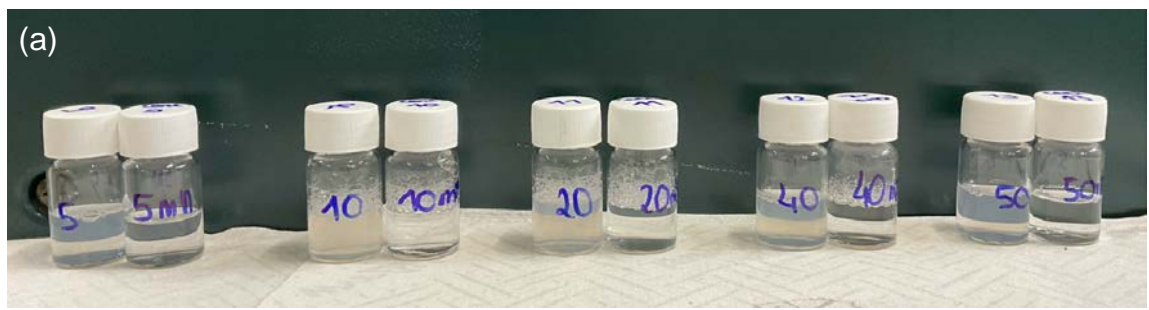

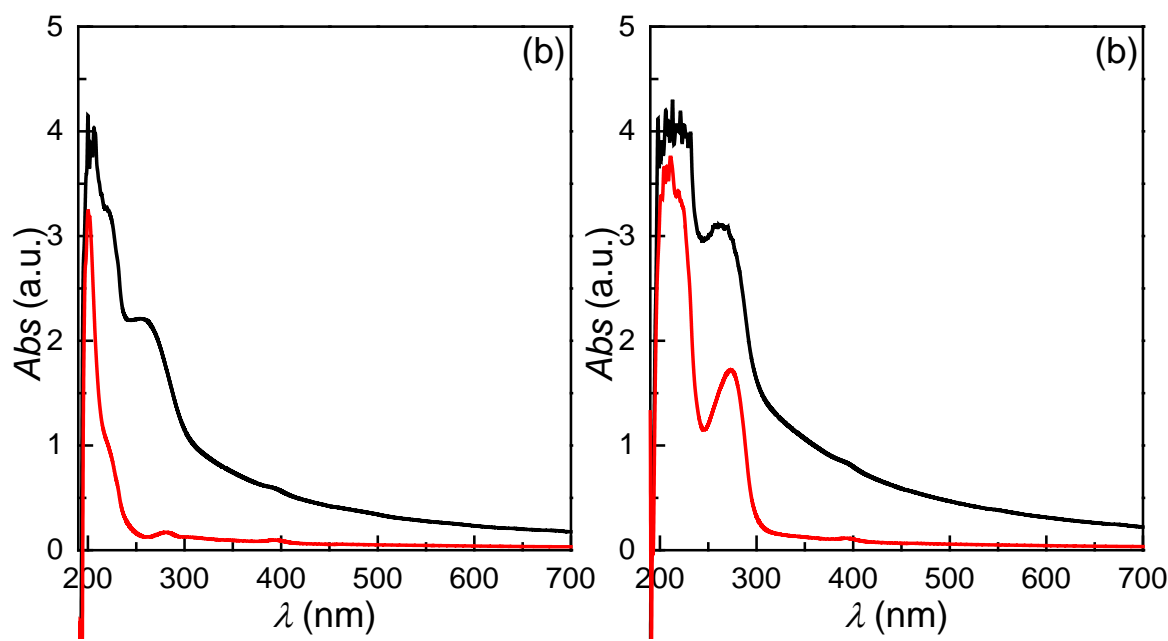

**Figure 5**. **(a)** Images showing CHI-APG mixtures with a fixed chitosan concentration (1 mM) and increasing APG concentrations (from left to right 5, 10, 20, 40 and 50 mM) at a NaCl concentration of 10 mM. For each sample two vials are shown, one corresponding to the sample as prepared (left vial) and a second one corresponding to the supernatant obtained after centrifugation (right vial). **(b,c)** UV spectra for CHI-APG mixtures with fixed chitosan concentration (1 mM) and NaCl (10 mM) concentrations and two different APG concentrations: 10 mM **(b)** and 20 mM **(c)**. Both panels show two spectra, one corresponding to the sample as prepared (black curve) and a second one corresponding to the supernatant obtained after centrifugation (red curve).

The UV-visible absorption spectra reveal marked differences between freshly prepared samples and their corresponding supernatants post-centrifugation. While the changes in band intensity can be directly linked to material depletion during centrifugation, which lowers the overall concentration and hence reduces absorbance, other changes in the spectra suggest more intricate phenomena. In the visible region, where no characteristic absorption bands are present, differences in absorbance may be attributed to turbidity variations. Fresh samples exhibit greater turbidity, resulting in increased light scattering and, consequently, higher absorbance values. Conversely, the supernatants obtained after centrifugation show reduced scattering, resulting in lower absorbance values. In the UV region, however, it can be observed more complex changes in both the intensity and shape of absorption bands, implying effects beyond a simple concentration reduction. These spectral alterations suggest the formation of distinct phases with differing compositions within the initial mixtures. If the system were homogeneous, centrifugation would produce a proportional decrease in absorbance across the spectrum. However, in the here obtained spectra, the direct comparison of the intensities may be affected by the fact that the spectra appear saturated in different points of the UV region. Despite the difficulty of discussing the influence of the depletion in the band intensities, the analysis of the band profile for the samples obtained and the obtained supernatant upon centrifugation suggests important differences. For instance, in Figure 5b, there is no evidence in the spectrum of the supernatant of the band appearing in the wavelength range 250-300 nm for the sample before centrifugation. The modification of the characteristic bands is also found in the analysis of the spectra in Figure 4c. Therefore, the presence of changes in band characteristics suggests the existence of a selective partitioning of components between the phases, with specific constituents concentrating in the dense, sedimented phase, while others preferentially remain in the supernatant. The selective partitioning underscores the heterogeneous nature of the system. The above observations point to a complex partitioning process in the system: components initially dispersed in the continuous phase selectively aggregate into a dense phase during centrifugation. This aggregation is likely driven by entropic effects favouring phase separation under specific conditions, leading to the observed sedimentation.

**3.2. Effective charge of Chitosan-APG mixtures**

The phase behavior of chitosan-APG system may be explained by considering that at low APG concentration, the binding of APG occurs randomly along the polymer chain with a negligible role of the interactions between the alkyl tails of different surfactant molecules. This leads to the formation of relatively hydrophilic complex characterized by a charge similar to that of chitosan chains as demonstrated by the zeta potential ($\zeta$ potential) measurements reported in Figure 6. However, when the APG concentration overcome the threshold value, the system enters in the coacervation region. In this region, the number of bound surfactant molecules to chitosan chains become high enough to consider the influence of van der Waals interactions between the alkyl tails of different APG molecules. Under these conditions, it may be expected the coacervate droplets adopts a core-shell structure where the inner core is formed by the hydrophobic tails of the surfactants surrounded by a charged hydrophilic shell where the charged monomers of CHI are placed, resulting in the increase of the $\zeta$ potential.

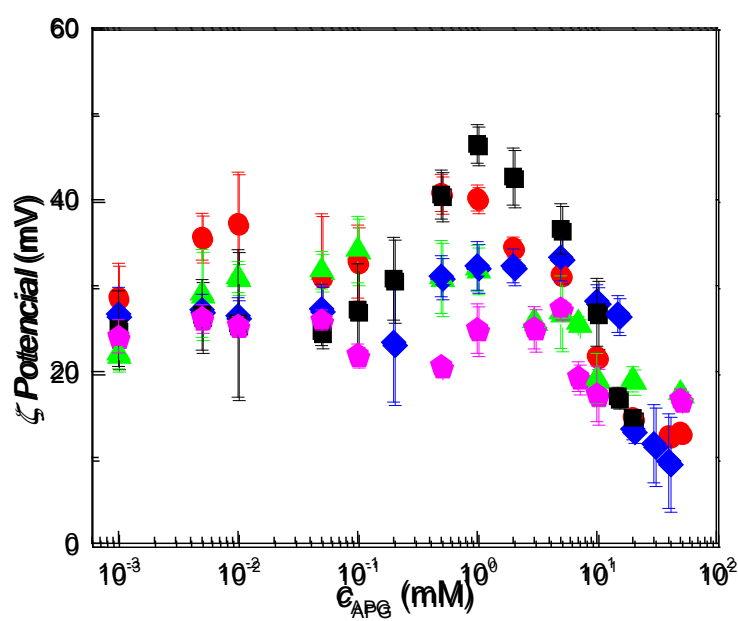

**Figure 6.** Dependence of the $\zeta$ potential on APG concentration for CHI-APG mixtures, with fixed 1 mM chitosan concentration, prepared at different NaCl concentrations ($c_{NaCl}$). Different NaCl concentrations are represented by different coloured symbols. 0 mM (■), 10 mM (●), 40 mM (♦), 100 mM (▲) and 150 mM (⬟).

The decrease in the $\zeta$ potential at the highest APG concentrations can be justified considering that the number of APG molecules in solution is extremely high, and at a certain point start to bind to the chitosan chain as micelles. This results in a redissolution of the coacervate droplets, which leads to the shielding of the charge of the chitosan chains. This pushes the effective surface charge of the aggregate to values close to the electroneutrality, similarly to that what happens for the interaction of silicon dioxide nanoparticles and copolymers belonging to the Pluronic family.[38] It is important to highlight that increasing the ionic strength of the solution leads to a reduction in the zeta potential values of the CHI-APG complexes. In this case, the electrostatic of the systems plays a very important role. In fact, the reduction in the effective charge of the chitosan chains can be explained in terms of the salt screening effect. As the concentration NaCl increases, the counterions from the salt interact with the charged groups on the chitosan chains, diminishing their net charge. Consequently, this reduction in the effective charge density of chitosan not only reduces the zeta potential of the individual polymer chains but also affects the overall charge of the CHI-APG aggregates.

**3.3. Adsorption of Chitosan-APG mixtures on negatively charged surfaces**

As it is often observed in oppositely charged polyelectrolyte-surfactant mixtures, a clear correlation can be established between the different states of the complexes in solution and their adsorption onto solid surfaces.[39–41] Figure 7 shows the acoustic thickness ($h_{ac}$), measured using QCM-D, as a function of surfactant concentration for the adsorption of polyelectrolyte-surfactant mixtures on a negatively charged surface.

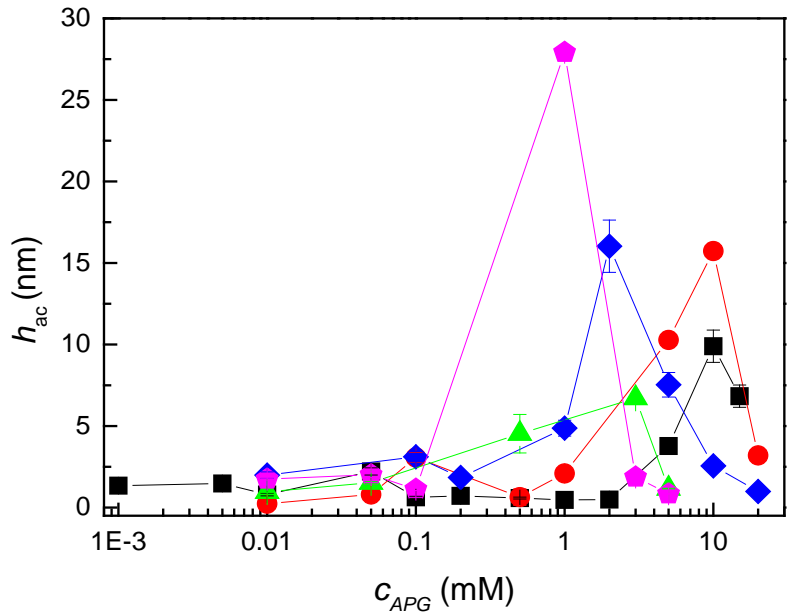

**Figure 7.** Dependence of the acoustic thickness on APG concentration for the deposition of CHI-APG mixtures, with fixed 1 mM chitosan concentration, prepared at different NaCl concentrations ($c_{NaCl}$) on negatively surfaces. Different NaCl concentrations are represented by different coloured symbols. 0 mM (■), 10 mM (●), 40 mM (♦), 100 mM (▲) and 150 mM (⬠).

The results indicate that, irrespectively of the ionic strength, the adsorbed amount, represented by the acoustic thickness, exhibits three distinct regions that closely mirror the phase behavior of the system. At low surfactant concentrations, the adsorption of polymer-surfactant complexes on the negatively charged surface resembles that observed for chitosan solutions alone, which agrees with the similar effective charge of chitosan and complexes. Then, when the system enters in the coacervation an enhanced deposition mediated by the coacervate droplets occurs. This enhanced deposition reaches a maximum for concentrations similar to that corresponding to the maximum in turbidity. Once the coacervation region is overcome, the deposition drops down to values close to zero. This is compatible with the reduced effective charge of the complexes in this region, which hinders its deposition on a negative charge surface. On the other hand, the enhancement of deposition becomes more pronounced as the ionic strength of the solution increases. This can be explained by considering the effect of ionic strength on the solubility of chitosan in water. As ionic strength increases, water becomes a poorer solvent for chitosan, reducing its solubility. This leads to a most favoured depletion of the aggregates from the solution due to the reduction in their effective charge, consistent with the zeta potential results. These findings suggest that the phase behavior of the system plays a critical role in governing adsorption.

## 4. Conclusions

This study demonstrates that coacervation-induced deposition can occur in polyelectrolyte-surfactant systems even in the absence of direct electrostatic interactions between the polyelectrolyte chains and surfactant molecules. The findings reveal that manipulating the ionic environment provides a tunable parameter for controlling coacervation, with higher ionic strengths enhancing hydrogen bonding and facilitating coacervate formation. This approach not only expands fundamental knowledge but also opens new perspectives in the design and formulation of complex fluids across diverse fields, such as drug delivery, personal care products, food technology, and enhanced oil recovery, where polyelectrolyte-surfactant systems are crucial.

Additionally, the study highlights the importance of hydrogen-bonding interactions in driving coacervation and controlling the structural properties of the resulting complexes, particularly in the formation of core-shell morphologies. This offers promising pathways for the development of targeted delivery systems and responsive materials where deposition characteristics can be modulated by adjusting ionic conditions. Future advancements require a deeper understanding of the performance of mixtures under conditions that resemble those existing under real operation conditions, making a more specific analysis of the role of salinity and salt type as well as of the surfactant concentration and dilution on the behavior of these mixtures.

**Acknowledgement**

This work was funded by the grant PID2023-147156NB-I00 of the MCIN/AEI/10.13039/501100011033 (Spain), the grant PR12/24-31566 (Ayudas para la Financiación de Proyectos de Investigación UCM 2023) and the European Innovative Training

Network-Marie Sklodowska-Curie Action NanoPaInt (grant agreement 955612) of the E.U. The authors express their gratitude to the Unidad de Espectroscopía y Correlación (CAI de Técnicas Químicas) at Universidad Complutense de Madrid for granting access to their facilities.

**Author contributions**

Ana Puente-Santamaría: data curation; formal analysis; investigation; methodology; software; visualization; writing-review and editing. Josselyn N. Molina-Basurto: investigation. Eva Gerardin: investigation. Francisco Ortega: funding acquisition investigation; project administration; resources; validation; writing-review and editing. Ramón G. Rubio: conceptualization; funding acquisition investigation; project administration; resources; supervision; writing-review and editing. Eduardo Guzmán: conceptualization; data curation; formal analysis; funding acquisition; investigation; project administration; software; resources; supervision; validation; visualization; writing-original draft.

**Conflicts of interest**

There are no conflicts to declare.

**Data availability**

Data are available upon reasonable request.